\documentclass[prb]{revtex4}    
\usepackage{times}
\usepackage{graphicx}
\usepackage{times}
\usepackage{graphicx}

\begin{document}

\title{Thermodynamics of DNA packaging inside a viral capsid:
the role of DNA intrinsic thickness}

\author{Davide Marenduzzo$^1$ and Cristian Micheletti$^2$\\
\small $^1$ Department of Physics, Oxford University, 1 Keble Road, Oxford OX1 3NP, United Kingdom\\
\small$^2$ International School for Advanced Studies (S.I.S.S.A.) and INFM, Via Beirut 2-4, 34014 Trieste, Italy}

\date{\today}

\begin{abstract}
We characterize the equilibrium thermodynamics of a thick polymer
confined in a spherical region of space.  This is used to gain insight
into the DNA packaging process.  The experimental reference system for
the present study is the recent characterization of the loading
process of the genome inside the $\phi$29 bacteriophage capsid.  Our
emphasis is on the modelling of double-stranded DNA as a flexible
thick polymer (tube) instead of a beads-and-springs chain.  By using
finite-size scaling to extrapolate our results to genome lengths
appropriate for $\phi$29, we find that the thickness-induced force may
account for up to half the one measured experimentally at high packing
densities.  An analogous agreement is found for the total work that
has to be spent in the packaging process. Remarkably, such agreement
can be obtained in the absence of any tunable parameters and is a mere
consequence of the DNA thickness. Furthermore, we provide a
quantitative estimate of how the persistence length of a polymer
depends on its thickness. The expression accounts for the significant
difference in the persistence lengths of single- and double-stranded
DNA (again with the sole input of their respective sections and
natural nucleotide/base-pair spacing).
\end{abstract}

\maketitle

\section{Introduction}

The goal of the present study is to characterize the equilibrium
thermodynamics of a thick polymer chain confined in a spherical
cavity. In particular we aim at establishing what force is necessary
to apply, and how much energy is to be spent, to achieve a given
packing density for the polymer. This problem is intimately related to
the problem of DNA packaging inside a viral capsid
\cite{phi29review,bloomfield,reviewcapsid}, which is under an
increasing theoretical and experimental attention
\cite{phi29,phi29motor,gelbart,odjik,pais,muthukumar,podgornik}. At
variance with previous studies we model explicitly the double-stranded
(ds) DNA as a cylindrical tube with a finite thickness (as opposed to
a beads-and-springs model) and study how it impacts on the force and
work required to package it. We shall further discuss how a finite
thickness induces a non-negligible persistence length in a polymer.
The experimental reference for the present study is provided by a
series of experiments \cite{phi29} where the loading process of DNA
inside the $\phi$29 virus head was followed under controlled
conditions. In this situation, the 6.6 $\mu$m long double-stranded DNA
polymer was fed by a molecular motor inside the virus head, a prolate
$54$ nm by $42$ nm icosahedron\cite{phi29,cell,bloomfield}.  By using
suitable force-feedback measurements it has been possible to carefully
measure the force exerted by the motor at various stages of the
loading process, and hence the total energy required for the whole
packaging process.  It has been found \cite{phi29} that the internal
force exerted by the packaged DNA against the $\phi 29$ capsid walls
(which has to be overcome by the portal motor), displays a dramatic
increase during the final loading stages.  In particular, the maximum
value attained by this internal force is $\sim 50$ pN. This is a
strikingly large number when compared to typical forces encountered at
the molecular and cellular scale.

In a recent study\cite{gelbart}, a molecular dynamics simulation has
been used to characterize the dynamical packaging of DNA in a
spherical cavity. In that approach, the double stranded DNA was
coarse-grained in a beads-and-strings chain subject to
self-interaction. {Interestingly, it was found that, at high
packing densities, the resistance to packaging was not dominated by
the DNA self-interaction (whether attractive or repulsive).  An
insight into this result can be obtained by a soluble mean-field
model, presented later, which highlights how the divergence of the
packing force at high density is not controlled by self-interactions.}

These results stimulate the search for an alternative physical
mechanism responsible for the onset of the large packaging pressure
encountered e.g. in $\phi$29.  For this reason, we adopted a minimal
model where DNA is treated as a thick chain of equispaced beads. {
Since the DNA self-attraction or repulsion does not appear to be
crucial at high packing densities, we have intentionally kept the
model at the simplest level and thus avoided the introduction of any
explicit treatment of the known self-interaction terms
\cite{strey}. The only two parameters entering in our model are the
spacing between the beads and the chain thickness; we do not attempt
any tuning of these parameters but, instead, adopt the corresponding
values desumed from the experimental literature. In particular, the
natural base spacing is set to 0.34 nm while the radius of hydrated
double-stranded DNA is taken as 1.25 nm \cite{bloomfield}.} Note that
in this work the thickness of a tube is defined as the radius of its
circular section taken normally with respect to the tube centerline.

As we shall show below, the appropriate treatment of the finite
thickness of the biopolymer allows to account for a variety of elastic
properties of DNA in a natural way,{even in the absence of
additional phenomenological parameters, such as the bending rigidity,
that are used in standard DNA coarse-grained models}. For example, it
will be shown that the mere difference in thickness of single-stranded
and double-stranded DNA can account for the large change in the
respective persistence lengths.  More importantly, by considering the
force necessary to compactify thick chains of increasing number of
beads, we estimate the packaging force (and work) which turns out to
be in good quantitative agreement with experimental measures for
$\phi$29 under the same density conditions. The present analysis
involves finite-size scaling techniques to extrapolate results
obtained in equilibrium stochastic simulations of chains with up to
200 beads. The most compact conformations obtained for such lengths
reveal the clear tendency of thick polymers to occupy the spherical
cavity by adopting a spool-like conformation, consistently with
previous expectations \cite{bloomfield,podgornik,odjik} and with the
recent numerical study of Kindt {\em et al.} with a different
model \cite{gelbart}.

\section{Theory and Methods}

Due to the large number of atomic constituents of both inorganic and
biological polymers, it is necessary to resort to suitably simplified
models in order to characterize their physical behaviour by means of
analytical or numerical studies. Customarily the polymeric chain is
coarse-grained into a succession of discrete beads. Besides capturing
the effect of chain connectivity through the introduction of suitable
strings or springs between consecutive beads, it is usually necessary
to introduce appropriate interactions between the beads in order to
capture the salient physical features of the given system. For the
case of ds DNA the latter would include effects such as strands
self-interaction (either attractive or repulsive) and bending
rigidity.  As a useful term of comparison for the present and previous
numerical studies of DNA packaging, it is useful to consider a
mean-field model (inspired by the treatment in
Refs.\cite{garel,lise}) which, while incorporating the above
mentioned features, is still amenable to an analytic treatment. This
model consists of a polymer chain of $N$ beads embedded in a discrete
three-dimensional space and subject to a contact interaction,
$\epsilon$, and a bending rigidity (corner penalty), $h$. {Within a
mean-field picture \cite{garel,lise} it is possible to obtain
explicit expressions for the free energy, $F_N$ and loading force,
$f$, required to package the chain inside a cube of side $L$:

\begin{eqnarray}
\label{1}
\frac{\beta F_N(x)}{N}=\frac{1-x}{x}\log{(1-x)}+\beta\epsilon(1-3x)\\
\nonumber -\log{\left[\frac{2+4\exp{(-\beta h)}}{e}\right]},\\ 
a\, \beta f(x)=
-\log{(1-x)}+\beta\epsilon(1-6 x)\\ \nonumber
-\log{\left[2+4\exp{(-\beta h)}\right]}
\label{eqn:force}
\end{eqnarray}

\noindent where $x$ is the density $x = {N/L^3}$, $\beta=(k_B T)^{-1}$
with $k_B$ the Boltzmann constant and $a$ is the unit length in the
system.  It is important to notice that the free-energy per bead,
$F_N/N$ depends on $N$ and $L$ only through the density $x$; this fact
will be exploited to analyze the results for our confined off-lattice
chains with finite thickness.

Despite the simplified nature of the mean field model, the approximate
force of expression (\ref{eqn:force}) provides a useful starting point
for both understanding recent packaging studies as well as suggesting
new improvements of traditional models.  The crucial observation is
that in the tight-packing limit, $x \to 1$, the divergence of $f(x)$
is entirely controlled by the logarithmic term which is independent of
both $\epsilon$ and $h$. This result provides a theoretical framework
for explaining the numerical findings of Ref.\cite{gelbart} who
ascertained that the force loading curve was not too sensitive to the
potential strength and sign at high packing densities.  In other
words, from this simple mean-field treatment we have an indication
that a resistance to dense confinement comes from entropic effects
rather then polymer self-interactions or flexibility.  This is also in
agreement with the picture proposed in Ref. \cite{strey}.}

To the best of our knowledge the fact that also the finite thickness
of DNA is responsible for its elastic properties has not been fully
investigated before, and hence is the focus of the present
study. Usually, in beads-and-springs models, excluded volume effects
are incorporated at the level of hard-core repulsion between pairs of
coarse-grained beads. However, the most satisfactory way to account
for steric effects is through the modelling of the polymer chain as a
tube with finite uniform thickness, $\Delta$, rather then a series of
beads connected by one-dimensional springs.
The appropriate treatment of the chain thickness usually leads to
remarkably different physics than with traditional beads-and-springs
models. A striking example is provided by the natural emergence of
protein-like secondary motifs in marginally-compact conformations of
thick chains\cite{opthelix}.

\begin{figure}[htbp]
\centerline{\includegraphics[width=2.0in]{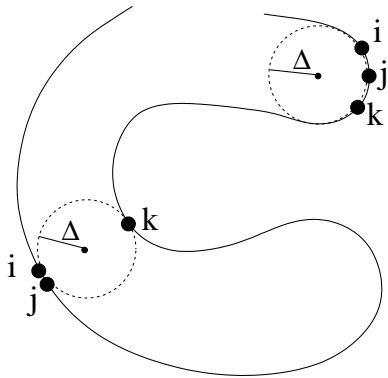}}
\caption{In order for a curve to be a viable centerline for a
tube of uniform thickness $\Delta$ it is necessary that the
radii, $r_{ijk}$ of the circles going through any triplet of points on
the curve $i,\ j\ k$ are not smaller than $\Delta$. This requirement
forbids the presence of tight local bends incompatible with the tube
thickness as well as unphysically proximity of distinct portions of the
centerline.}
\label{fig:0}
\end{figure}

The finite thickness impacts on two distinct features of polymer
conformations. On one hand it will constrain the local radius of
curvature to be not less than $\Delta$ to avoid singularities.  This
effect induces naturally a bending rigidity on the polymer. As will be
shown later, the resulting persistence length depends crucially on
$\Delta$, so that the latter appears to be a decisive factor for the
huge difference of the persistence length for single-stranded and
double stranded DNA, as we shall show later.  On the other hand there
is also a non-local effect due to the fact that any two portions of
the tube, at a finite arclength separation, cannot interpenetrate.
Equivalently, the centerlines of the two portions need to be at a
distance greater than 2$\Delta$ (see Fig. \ref{fig:0}).

In traditional beads-and-springs models it is only this second effect
that is taken into account through pairwise hard-core
repulsion. Interestingly, one needs to go beyond pairwise interactions
to account for the above mentioned effects in discretised polymer
chains. This is best illustrated considering the discrete succession
of equispaced "beads", ($\vec{r}_1, \vec{r}_2,... \vec{r}_N$), as the
centerline of a tube with thickness $\Delta$. The requirement on the
local radius of curvature is simply enforced by finding the radii of
the circles going through any consecutive triplet of points and
ensuring that each of them is greater than $\Delta$.  The non-local
effect can be addressed within the same framework by considering the
minimum radius among circles going through any non-consecutive triplet
of points. This radius is precisely the distance of minimum approach
and, again, has to be greater than $\Delta$.  In summary, the
appropriate treatment for the finite thickness, $\Delta$, of the tube
associated to a discretized centerline is reflected in the fact that
the radii of the circles going through any triplet of distinct points
has to be greater than $\Delta$\cite{Thickness1}.

One is thus naturally led to consider the following Hamiltonian for
the unconstrained, non self-interacting, double-stranded DNA:

\begin{equation}
 {\cal H}(\Gamma) = \sum_{ijk} V_3(r_{ijk})
\label{eqn:ham}
\end{equation}

\noindent \noindent where $V_3$ is the three-body potential used to
enforce the thickness $\Delta$ of the chain
\cite{Thickness1,Thickness2,jstatphys,opthelix,Thickness3,secstr}
and $\Gamma$ is the configuration of the ds DNA. 
The argument
of $V_3$ is the radius of the circle going through the triplet of
distinct points $i,\ j,\ k$ and has the form
\begin{equation}
V_3(r) = \left\{ 
\begin{array}{l l}
0 & \mbox{if $r > \Delta$,} \\
+\infty & \mbox{otherwise.} 
\end{array}
\right .
\end{equation}

\noindent The tube ends were modelled as hemispheres by introducing
two ``phantom'' beads essentially coinciding with the terminal beads of
the tube.

Since our experimental reference case is the packaging of genome
inside the $\phi$29 capsid, we take $\Delta$ as the thickness
appropriate for hydrated double-stranded DNA: $1.25$ nm while the
spacing between the beads of our discretized tube is naturally taken
as the average base spacing: $a=$0.34 nm.

As stated earlier our goal is to describe the equilibrium
thermodynamics of a self-avoiding thick chain of $N$ beads confined
in a sphere of radius $R$. In particular, we aim at calculating the
free energy $F_N(R)$, for which it is straightforward to calculate the
work done to package it under conditions of no energy dissipation,
and also the quantity:

\begin{equation}
\Delta G_N(R) = F_N(R) - F_N(\infty)\ .
\end{equation}

\noindent The force that is necessary to apply in order to
feed additional chain beads inside the sphere can be calculated by
differentiating $\Delta G_N(R)$ with respect to $N$:

\begin{equation}
f \propto {\partial \Delta G_N(R) \over \partial N} \ .
\end{equation}

In principle, the free energy, $F_N(R)$ could be obtained through a
Monte Carlo simulation where the conformation space is restricted to
only those structures that can be confined in a sphere of radius
not larger than $R$. This is, however, impractical due to the fact
that in situations of high density (small $R$) it will be very
difficult to move through distinct compact structures while respecting
the confining constraints.

A useful alternative is to work in the conjugated ensemble where,
instead of keeping fixed the radius of the confining sphere, $R$, one
applies a uniform ``hydrostatic'' pressure, $P$ that compactifies the
chain. In this situation, the Boltzmann weight of a configuration,
$\Gamma$ is given by:
\begin{equation}
e^{- (E + PV)/K_B\,T}
\end{equation}

\noindent where the temperature $T$ is taken as the room temperature,
$T=300 \, K$, $E$ is the structure energy calculated via Eq.
(\ref{eqn:ham}) and $V$ is the volume of the spherical hull associated
to $\Gamma$. To find the radius of the spherical hull enclosing
$\Gamma$ we first determine the maximum distance, $\bar{r}$ from the
centre of mass of any of the beads in $\Gamma$. Since these beads lie
on the centerline of a tube of thickness $\Delta$, the required radius
is $R = \bar{r}+\Delta$. It should be noted that, in principle, the
smallest confining sphere is not necessarily centred on the centre of
mass of $\Gamma$. However, we have verified by an explicit numerical
search of the location of the centre of the smallest sphere that this
approximation is very good when dealing with structures with good
overall compactness, which are the focus of this investigation. We
stress that $E$ can take only two values, 0, and $\infty$ for
configurations that are respectively compatible or incompatible with
the thickness requirement. This effectively restricts structure space
to the ensemble of viable conformations of the thick tube.

By working at fixed temperature, one can sample the configuration
space for various values of $P$ by using a standard Metropolis
criterion. From the distribution (histogram) of the various hull
volumes encountered at various pressures one can use an ordinary
multiple histogram technique \cite{ferren} to recover the density of
states, $W(R)$ corresponding to the number of viable structures
(i.e. compatible with the preassigned thickness $\Delta$) and hull
radius $R$.

The required free energy, $F(R)$ is then:
\begin{equation}
e^{- \beta F(R)} \propto \sum_{R^\prime < R} W(R^\prime) 
\end{equation}

The reconstructed free energy, $F(R)$, is defined up to an additive
constant, which reflects the fact that the density of states can be
obtained only up to a multiplicative constant. This additive constant
is immaterial for the calculation of the packaging work and force in a
system with a given number of beads, though it has to be set
appropriately in order to simplify the scaling analysis of $F_N(R)$ as
a function of $N$ at fixed $\beta$.  For this reason we have taken the
unconstrained situation, $R = \infty$, as the reference case
associated with zero free energy: $F_N(R = \infty) =0$ (see Eq. 5).

It is worth to point out that the force estimate in (\ref{eqn:force})
is for a chain that is fully confined in the cube, i.e. we have not
considered the case of a chain that is partially contained in the cube
while the remainder constitutes an external tail. The latter case may
be closer to the {\it in vivo} condition; however, the
contribution of the floating tail to the free energy is expected to
become less and less important compared to that of the confined
portion as $x \to 1$ in Eqs. \ref{1} and 2, 
which is the limit case of interest. In
addition, in experiments where the packaging force is probed, the
protruding tail is kept under considerable mechanical tension and
therefore is expected not to make significant contribution to the
system entropy.

\section{Results}

One of the paradigm models fruitfully used to describe DNA chains is
provided by the freely-rotating chain \cite{Daune}. Within this scheme
a DNA chain is described as a succession of equally-long bond vectors,
$\vec{t}_i$, subject to the constraint that consecutive bonds must be
at a given angle, $\alpha$. This constraint directly impacts on the
chain persistence length (which measures how correlations between two
bond vectors decrease with their arclength separation).

For a given length of the bond vectors, the bonding angle $\alpha$ has
to be set {\em a posteriori} so to reproduce e.g. the experimental
persistence length, $\xi$. It is appealing to notice, as pointed out
before, that the notion of intrinsic thickness of the DNA chain,
naturally impacts on the minimum attainable radius of local curvature
and hence on the allowed range of the bonding angle, $\alpha$. This
leads to the existence of a non-trivial persistence length, $\xi$,
induced by the chain thickness. Such dependence can be estimated as
follows.  Consider a discrete chain of thickness $\Delta$, with the
consecutive unitary bond vectors $\vec t_i\equiv \vec r_{i+1}-\vec
r_i$ and $\vec t_{i+1}$.  The constraint over the bonding angle is
give by

\begin{eqnarray}
\vec t_i\cdot\vec t_{i+1} & \ge & 1-\frac{1}{2\Delta^2}  \ .
\label{eqn:constraint_thickness} 
\end{eqnarray}

\noindent Neglecting the effects of the chain self-interactions it is
possible to calculate explicitly the quantity $\langle \vec
t_n\cdot\vec t_m\rangle$, with $n<m$, which defines in turn a
persistence length, $\xi$ via $\vec t_n\cdot\vec t_m\sim
\exp{\left(-\frac{|n-m|}{\xi}\right)}$. One thus obtains that
$\xi(\Delta)$ is given by

\begin{equation}\label{persistence_thick_polymer} 
\xi(\Delta)=-\frac{1}{\log{\left(1-(2\Delta)^{-2}\right)}}\ . 
\end{equation}

\begin{figure}[tbp]
\centerline{\includegraphics[width=3.0in]{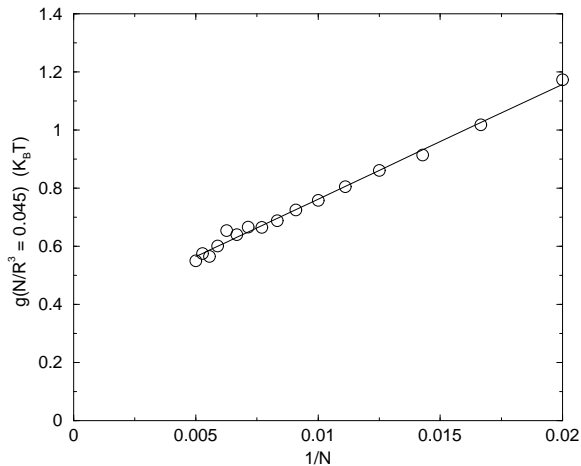}}
\caption{Curves for the reduced free energy  at the fixed 
density $N/R^3= 0.045$ for values of $N$ in the range $50 \le N \le 200$.}
\label{fig:b}
\end{figure}

This approximate relationship can be specialized to the case of
double-stranded DNA, where the unit bonds are taken as the base
spacing, 0.34 nm and the thickness equal to the radius of hydrated
dsDNA, 1.25 nm. One thus observes $\xi= 19$nm, which is about one half
of the correct experimental value $\xi_{exp} \approx 50$ nm. This
theoretical estimate, which takes into account the mere effect of
steric hindrance of the chain does not include the contribution of
electrostatic repulsion of the phosphate groups to the apparent
persistence length\cite{cozzarelli} but appears to capture the
correct order of magnitude of the observed persistence length.

As a further validation of this approach it is worth considering the
case of single-stranded DNA. In this case the vector length is equal
to $\sim$ 5 \AA, which is the typical nucleotide separation projected on a
line interpolating the chain, while the thickness is about $\sim$ 4
\AA. The near equality of these two parameters imply that the
thickness induced persistence length of single-stranded DNA is
comparable with the nucleotide spacing, and in fact one obtains, $\xi$
$\sim$ 1 nm.  Again, this is consistent with the experimental data for
ssDNA in solutions of high ionic strength (which screen the phosphate
charges) for which values for $\xi$ of the order of 1 nm are observed
\cite{plengthssdna1}.

Given the simplified derivation of the parametric dependence of $\xi$
on $\Delta$ and the total absence of adjustable parameters in the
theory, this agreement is remarkable and testifies the need to model
explicitly the DNA intrinsic diameter.

While the soluble models leading to e.g. the results of eqns
(\ref{eqn:force}) and (\ref{persistence_thick_polymer}) are useful for
gaining insight in the basic physics of thick polymers, they cannot be
used to fully characterize more complicated instances such as the
packaging process. To address this issue it is, therefore, necessary
to resort to stochastic numerical simulations.  In the present study,
the equilibrium properties of the thick polymer was characterised by a
multiple Markov chain simulation\cite{TROW} where several copies of
the system each at a different pressure were evolved
simultaneously. The evolution was controlled by the Metropolis
acceptance of elementary chain distortion involving crankshaft, pivot
and slithering moves\cite{sokal}. A preliminary measure of correlation
times allowed to collect statistically independent measures for
reconstructing the density of states as described in the methods
section. The full equilibrium thermodynamics was thus obtained for
tubes with a discrete number of beads, $N$. Values of $N$ up to 200
were considered here. As was argued before, in the limit of large $N$
the free energy should be extensive in $N$ at fixed filling fraction,
$N/R^3$.  Therefore we expect $F_N(R)$ to obey the following scaling
relations:

\begin{equation}
\lim_{N \to \infty} F_N(R) \sim N \ g\left({N \over R^3}\right) \ ,
\label{eqn:g}
\end{equation}

\noindent which enables, for large $N$, to express the confining force
as a function of $g$:

\begin{equation}
a\, f_N(R)  = {\partial F \over \partial N} \sim  g\left({N
\over R^3}\right) + {N \over R^3} g^\prime\left(x\right)|_{x = N/R^3}\ .
\label{eqn:f}
\end{equation}

\noindent where $a=0.34$ nm is the unit length in our model. Note that
the same scaling theory applies also to expressions (\ref{1}) and
(\ref{eqn:force}).  These relations will be obeyed only approximately
for any finite chain, though such corrections decay for
increasing values of $N$, as shown in the following.

\begin{figure}[htbp]
\centerline{\includegraphics[width=3.0in]{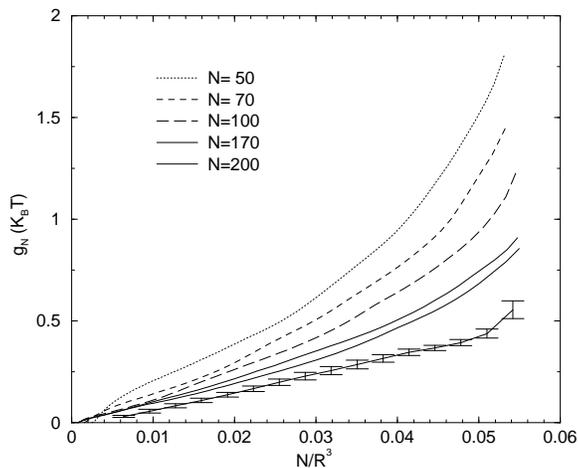}}
\caption{The reduced free energy as a function of the packing density,
$N/R^3$ for the indicated values of $N$. The bottom curve is obtained
by extrapolating the finite-size curves for $N \to \infty$. }
\label{fig:a}
\end{figure}

In Fig. \ref{fig:a} we have plotted the reduced free energy, $g_N =
F_N(R)/N$ as a function of $N/R^3$ for different values of $N$.  It is
apparent that, as the size of the system, $N$, increases, the curves
for $g_N$ deviate less from each other, indicating the decrease of
finite-size corrections from the limit curve of expression
(\ref{eqn:g}). Indeed, the systematic decay of such corrections for
increasing $N$ is visible in Fig. \ref{fig:b} where we have reported
the trend of $g_N$ as a function of $N^{-1}$ at a fixed density. As
visible in Fig. 3, the data are well interpolated by a linear
regression with correlation coefficient $r \sim 0.98$ . Through such
regression one can estimate the error on the extrapolated value for
$g$ as the error on the intercept.  The observed correlation
coefficient is statistically significant, since it pertains to 16 data
points and is suggestive that finite-size corrections decay
approximately as $N^{-1}$.  However, from the systematic decrease of
the linear regression slope observed when fitting data for the larger
tube lengths only, it can be argued that the convergence is faster
than $1/N$ and may also be compatible with decays such as $N^{-1.5}$.

The same rate of convergence of finite-size corrections is inherited
by the force obtained from (\ref{eqn:f}) by using $g_N$ instead of
$g$.  By considering several values of $N$ one obtains a succession of
curves for the loading force that, again, appear to approach a
limiting curve from above. This is in agreement with the physical
intuition that more force is required to pack a shorter (but equally
thick) chain to the same target density.

As for the corrections to $g$, also the force curves appear to
approach the limit curve of interest at a rate somewhat faster than
$1/N$. Nevertheless, since we cannot give a precise account of the
correction exponent we calculated the limit force by using the
conservative $1/N$ decay estimate and by fitting only the data of the
largest tube lengths, $100 < N \le 200$. The results are provided in
Fig. \ref{fig:d} and can be compared with the experimental curve of
ref. \cite{phi29} as discussed below.

If one considers the prolate $\phi$29 capsid as an ellipsoid the
reference density at full genome packaging as $x = 0.063$. The largest
densities reached in our simulation correspond therefore to about 85
\% of this value. Neglecting the contribution of the protruding tail
in the $\phi29$ experiment one can then compare the experimental force
at 85 \% genome packaging, 34 pN, with our peak force of 16 pN, which
appear to be correct within only a factor ot two. We stress that this
finding exploits only the severe restrictions in configuration space
operated by the finite thickness of DNA itself; it is striking that
this entropic effect appears to account for half the force observed in
the experiment. Furthermore, the limit curve for the force has been
obtained with the conservative $1/N$ estimate of the decay of
finite-size corrections; this arguably leads to an underestimation of
the force required to package the DNA chain.

\begin{figure}[tbp]
\centerline{\includegraphics[width=3.0in]{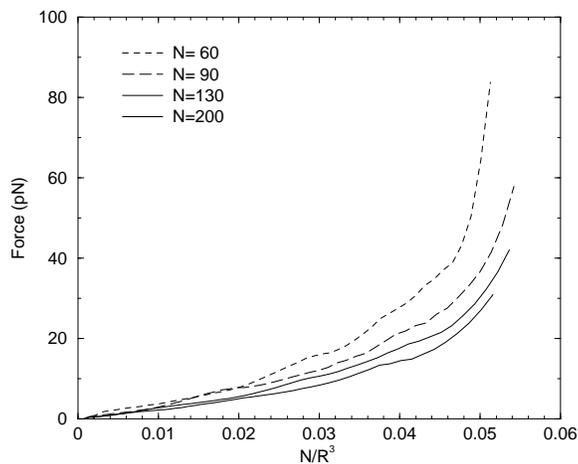}}
\caption{Curves for the force necessary to apply to package the thick
tube to the required density, $N/R^3$. The curves for the force are
obtained from simulations with the indicated number of beads, $N$.}
\label{fig:c}
\end{figure}

Another quantitative comparison against the experiment can be done
considering the total packaging work. The experimental total packaging
work in the experiment of ref \cite{phi29} was estimated as 2 $10^4\
K_B T$, corresponding to $1.04\ K_B T$ per base pair (or bead in our
model).  In our case, the work per bead necessary to package the tube
to a preassigned density, $x$, (starting from an unconstrained
situation) is given simply by $g(x)$ (see Fig. \ref{fig:b}). As
before, we can connect the peak density reached in our simulation with
the experimental situation where 85 \% genome is packaged in the
capsid. The value of $g$ observed in our model at this density
corresponds to 0.55 $K_B T$, which differs by less than a factor of
two from the experimental work (but the latter refers to 100 \%
packaging). In summary, even without accounting for the different
final densities in the experiment and in our model, we can
nevertheless conclude that also a substantial fraction (more than
50\%) of the work required to package DNA could result from the
entropic cost of compactifying a polymer of finite thickness.  These
results are also consistent, {\em a posteriori}, with the findings of
Ref. \cite{gelbart}, where it was pointed out that the detailed form
(and even sign) of the DNA self-interaction was not the determinant
factor in the packaging process.  The important role played by the
loss of configurational entropy in DNA packaging or condensation is
also supported by recent numerical and analytical work
\cite{harvey,odijk93}.

\begin{figure}[bp]
\centerline{\includegraphics[width=3.0in]{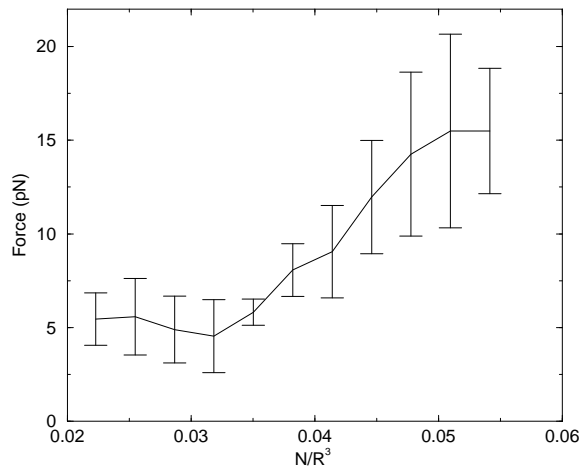}}
\caption{Plot of the extrapolated force required to package an
infinitely long tube to the preassigned density, $N/R^3$. The force is
expressed in picoNewtons. The estimated packing density of the fully-loaded
$\phi$29 capsid corresponds to 0.063 .}
\label{fig:d}
\end{figure}

As we argued before, the finite DNA thickness appears to account for a
significant fraction of the physical properties (such as persistence
length and resistance to confinement) that are usually ascribed
entirely to DNA self-interaction and bending rigidity.  As an
extension of the model discussed so far, one is lead to consider
the case where, on top of the finite-thickness treatment one adds
bending rigidity effects. The latter are customarily taken into
account by the following energy term:

\begin{equation}
E_{b.r.} = { K_B \, T \over 2} \xi \int \, ds {1\over R(s)^2}
\label{eqn:br}
\end{equation}

\noindent where the integral runs over the DNA contour, $R$ denotes
the local curvature and $\xi$ is the persistence length. In a model
which already incorporates DNA thickness the use of the full
persistence length in expression (\ref{eqn:br}) certainly
overemphasizes the elastic effects (since thickness alone can
satisfactorily account for a significant fraction of the DNA
persistence length, $\xi$). Even in this situation it will be apparent
that the additional (overestimated) force increment ascribable to the
presence of bending rigidity is of the same order of the one induced
by the thickness.

As a first step we provide an upper bound for the confining force
increment of the due to the added elastic term. To this purpose let us
consider a thick chain that has been compactified so to nearly fill a
given cavity. It is assumed that packing conditions are such that
there is almost no structural freedom left for how to add an
additional bead inside the hull. In particular, the bead has to be
added so to achieve the tightest local curvature of the chain. Since
the lower bound for the curvature is the thickness itself, $\Delta$,
the additional packaging force is estimated as:

\begin{equation}
\Delta f_{b.r.} = { K_B \, T \, \xi \over 2 \Delta^2}\ .
\end{equation}

\noindent Using $\xi = 50 nm$ as in Ref. \cite{gelbart} and $T =
300\, K$ one obtains that an upper bound for the force increment due
to bending rigidity is $\Delta f_{b.r.} \approx 66$ pN. To obtain a
more accurate estimate of $\Delta f$ we have undertaken additional
studies where the numerical scheme previously discussed for the
confined thick chain was generalised to include explicitly the bending
rigidity term of eq. (\ref{eqn:br}).

The augmented difficulty of packaging the chain in this situation
resulted in the fact that the maximum density achieved was $N/R^3=
0.050$, corresponding to 80 \% of the estimated packing fraction of
$\phi29$. In this case, the confining force for a chain of 200 beads
resulted equal to 39 pN. This corresponds to an extra force of around
$12$ pN, (the pure thickness-induced force at the same filling
fraction is $\sim$ $27$ pN, see Fig. \ref{fig:c}). 
This value overestimates the bending rigidity
contribution in the thermodynamic limit due to the systematic decrease
of the force as a function of chain length. However, even allowing
for the overemphasized contribution of bending rigidity due to both
finite-size effects and the use of the full pesistence length $\xi$ in
eq. (\ref{eqn:br}) one can already conclude that the finite-thickness
effects are comparable with those associated with bending rigidity
(see. Figs. 4 and 5).

This conclusion remains true when finite-size extrapolation is carried
out on this second model using chains of 40, 80,100,120,160 and 200
beads. From the thermodynamic data a total force of $24\pm 7$ pN can
be extrapolated at the highest densities $N/R^3= 0.050$. Despite the
still simplistic nature of the second model considered here, the
extrapolated force appears to be fully consistent with the
experimental results for phi29 at the corresponding filling fraction
of 80 \%, where a total force of 27 pN was recorded. Thus, the
different calculations presented here all indicate that the
contribution to the resistance to packaging arising from the finite
DNA thickness is comparable to that of other terms usually considered
in coarse-grained models of DNA. This indicates that various
phenomenological properties of DNA can be accounted in a more accurate
and complete way by including an explicit modelling of DNA thickness
besides the traditional terms in the effective Hamiltonian.

\begin{figure}[htbp]
\centerline{\includegraphics[width=3.0in]{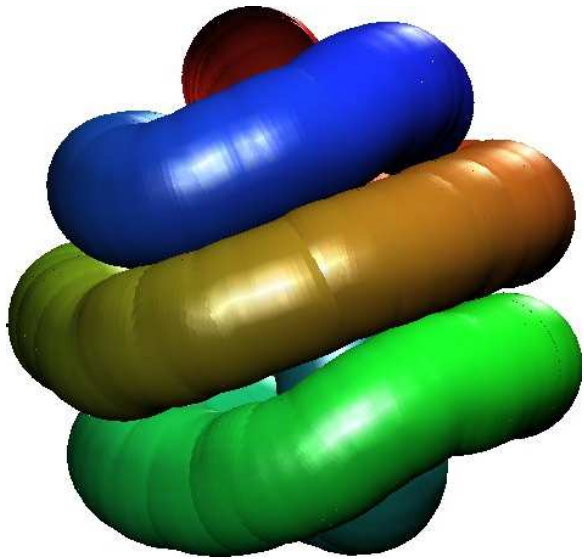}}
\caption{Example of a highly compressed conformation for a discrete tube of 200
beads and thickness $\Delta=3.7$ (in units of the bead spacing).}
\label{fig:f}
\end{figure}

Finally, it is interesting that, consistently with the study of
ref. \cite{gelbart} and with other theoretical
studies\cite{odjik,harvey} the most compact conformations obtained in
our stochastic simulations (in thermodynamic equilibrium) show the
marked tendency to occupy the spherical cavity by performing a
succession of turns in a spool-like fashion with a portion of the
tube, while the remainder of the chain fills the space around the
spool axis, as visible in Fig. \ref{fig:f}.

\section{Conclusions}

We have characterized the equilibrium thermodynamics of a polymer
chain with finite thickness confined inside a spherical cavity. The
model aims at schematizing the packaging of the genome
(double-stranded DNA) inside a bacteriophage capsid.  Stochastic
simulations and reweighting techniques were used to calculate the
force (and work) required to compactify to a given density a discrete
tube of up to 200 beads. The sole inputs of the model are the
experimental values for the spacing of the base pairs in
double-stranded DNA, and the thickness of the latter; no tunable
parameters or specific forms of self-interaction of the polymer are
introduced. By using finite-size scaling techniques we have estimated
the force necessary to package the genome inside the $\phi$29
virus. The predicted force (and packaging work) agrees with the
experimental one to within a factor of two, thus testifying the
importance of an appropriate modelling of the DNA thickness in the
packaging process. This fact is further corroborated by a theoretical
calculation which shows that the different thickness of single- and
double-stranded DNA can satisfactorily account for the large
difference of the respective persistence lengths.

ACKNOWLEDGMENTS We are grateful to S. C. Harvey, T. Odijk and
A. Maritan, for illuminating discussions and suggestions. We
acknowledge financial support from INFM and Murst Cofin 2001.

\end{document}